\documentclass{article}
\usepackage{ijcai25}
\usepackage{xcolor}

\usepackage{times}
\usepackage{soul}
\usepackage{url}
\usepackage[hidelinks]{hyperref}
\usepackage[utf8]{inputenc}
\usepackage[small]{caption}
\usepackage{graphicx}
\usepackage{amsmath,amsthm,amstext,amssymb}
\usepackage{mathtools}
\usepackage{booktabs}
\usepackage{algorithm}
\usepackage{algorithmic}
\usepackage{xspace}
\usepackage{comment}
\usepackage{multirow} 
\usepackage{float}
\usepackage[switch]{lineno}
\newcommand*{\Comb}[2]{\binom{#1}{#2}}

\newcommand\MC{\textsc{Math\-Check}\xspace}
\newcommand\CaDiCaL{\textsc{CaDiCaL}\xspace}
\newcommand\nauty{\textsc{nauty}}
\newcommand\AMS{\textsc{Alpha\-Maple\-SAT}\xspace}

\newcommand{\citein}[1]{\citeauthor{#1}~(\citeyear{#1})}

\newtheorem{lemma}{Lemma}

\title{Verified Certificates via SAT and Computer Algebra Systems for the \\ Ramsey $\boldsymbol{R(3,8)}$ and $\boldsymbol{R(3,9)}$ Problems}

\author{
Zhengyu Li$^1$\and
Conor Duggan$^2$\and
Curtis Bright$^3$\And
Vijay Ganesh$^1$\\
\affiliations
$^1$Georgia Institute of Technology, USA\\
$^2$University of Waterloo, Canada\\
$^3$University of Windsor, Canada
\emails
brian.li@gatech.edu,
c4duggan@uwaterloo.ca,
cbright@uwindsor.ca,
vganesh45@gatech.edu
}

\begin{document}

\maketitle

\begin{abstract}
The Ramsey problem $R(3,k)$ seeks to determine the smallest value of $n$ such that any red/blue edge coloring of the complete graph on $n$ vertices must either contain a blue triangle (3-clique) or a red clique of size~$k$. 
Despite its significance, many previous computational results for the Ramsey $R(3,k)$ problem such as $R(3,8)$ and $R(3,9)$ lack formal verification. 
To address this issue, we use the software \textsc{MathCheck} to generate certificates for Ramsey problems $R(3,8)$ and $R(3,9)$ (and symmetrically $R(8,3)$ and $R(9,3)$) by integrating a Boolean satisfiability (SAT) solver with a computer algebra system (CAS)\@. 
Our SAT+CAS approach significantly outperforms traditional SAT-only methods, demonstrating an improvement of several orders of magnitude in runtime. For instance, our SAT+CAS approach solves $R(3,8)$ (resp., $R(8,3)$) sequentially in 59 hours (resp., in 11 hours), while a SAT-only approach using state-of-the-art \CaDiCaL solver times out after 7 days. Additionally, in order to be able to scale to harder Ramsey problems $R(3,9)$ and $R(9,3)$ we further optimized our SAT+CAS tool using a parallelized cube-and-conquer approach. Our results provide the first independently verifiable certificates for these Ramsey numbers, ensuring both correctness and completeness of the exhaustive search process of our SAT+CAS tool.
\end{abstract}

\section{Introduction}

Ramsey Theory studies the existence of substructures with order within sufficiently large structures. The classical Ramsey $R(3,k)$ problem seeks the smallest integer $n$ (known as the Ramsey number) such that any red/blue edge coloring of the complete graph on $n$ vertices must contain either a blue triangle or a red $k$-clique. This problem is often framed as the ``party problem'': determining the minimum number of guests required to ensure that either three guests all know each other or $k$ guests are mutual strangers. Despite its simple formulation, computing Ramsey numbers is notoriously difficult, with only nine non-trivial values known to date, despite extensive research.

Most contemporary methods of finding non-trivial Ramsey numbers rely heavily on the use of computer programs such as \nauty~\cite{nauty2014} to enumerate graphs exhaustively. However, \nauty\ cannot generate a formal certificate certifying that the enumeration is indeed exhaustive, thus raising the possibility that these results may have errors in them. In this paper, we apply \MC~\cite{zulkoski2015mathcheck} and combine satisfiability solvers with computer algebra systems (SAT+CAS)~\cite{Bright2022} to not only produce formal certificates for Ramsey numbers but also achieve significant speedups compared to SAT-only approaches. A key component of our implementation is the use of the IPASIR-UP interface~\cite{fazekas2023ipasir} that facilitates seamless integration of external learned clause addition within modern SAT solvers. Specifically, we leverage IPASIR-UP to implement the {\CaDiCaL}+CAS framework as part of \MC. 
In this work, we focus on providing certificates of completeness for the exhaustive search components required to determine $R(3, k)$ and $R(k,3)$. It is important to emphasize that our approach does not constitute a formal proof of the Ramsey number. Instead, our certificates ensure correctness and completeness for the parts of the proof involving exhaustive graph enumeration. To follow the established convention in Ramsey theory, this paper begins by discussing the Ramsey number \( R(3, k) \). In later sections we find \( R(k, 3) \) aligns better with our computational approach, 
but \( R(3, k) = R(k, 3) \), as these are symmetric by definition.

\subsection{Our Contributions}

\paragraph{1) Certified Ramsey numbers:} In this paper, we extend the well-known SAT+CAS tool \MC~\cite{zulkoski2015mathcheck} (see Figure~\ref{fig:pipeline}) to solve Ramsey problems of type $R(3,k)$ and $R(k,3)$, and provide certificates of correctness.\footnote{All results can be reproduced using the public code from \url{https://github.com/ConDug/MathCheckRamsey}.} These certificates ensure the correctness of the exhaustive search process, though we cannot claim complete formal proofs since the SAT encodings have not been formally verified. We verify the values of two Ramsey numbers for $k= 8$~\cite{mckay1992value} and $k= 9$~\cite{grinstead1982ramsey}, whose proofs rely heavily on graph enumeration. To our knowledge, these are the only two known Ramsey numbers that have not been verified with proof certificates.

\paragraph{2) Speedup over SAT-only approaches:} We show that our sequential and parallel SAT+CAS tools significantly outperform SAT-only approaches (see Table~\ref{tbl:sec6}). This speedup is achieved through the isomorph-free exhaustive generation technique employed by \MC, which reduces the search space by dynamically blocking symmetric branches. 

\paragraph{3) Parallel cube-and-conquer SAT+CAS \MC:} We introduce a parallel cube-and-conquer SAT+CAS tool by integrating \AMS~\cite{alphamaplesat} as the cubing solver and \MC as the conquering solver. We solved and certified $R(8,3) = 28$ in 8 hours of wall clock time, and in doing so
decreased the total CPU time spent solving to 6.2 hours (while the sequential version spent 18.5 hours just solving), and solved and verified $R(9,3) = 36$ in 26 hours (while the sequential version timed out after 7 days).

\section{Preliminaries}

\subsection{SAT+CAS for Combinatorial Problems}

A conflict-driven clause learning (CDCL)~\cite{cdcl1,cdcl2,cdcl3} satisfiability (SAT) solver takes as input a Boolean formula in conjunctive normal form (CNF) and determines whether there exists an assignment of variables making the formula evaluate to true, in which case the formula is satisfiable (SAT)\@. 
Otherwise, the formula is unsatisfiable (UNSAT)\@.
Thanks to the rich advancement made by the SAT community, state-of-the-art CDCL SAT solvers can solve many instances with millions of variables efficiently.
However, SAT solvers face challenges when solving hard combinatorial problems such as the Ramsey problem due to the large amount of symmetry in the search space.

Computer Algebra Systems (CASs)
like Maple~\cite{maple}, Mathematica~\cite{Mathematica}, Magma~\cite{magma}, and SageMath~\cite{sagemath}
are storehouses of mathematical knowledge, containing state-of-the-art algorithms from many mathematical areas.
Therefore, many mathematical constraints can be easily expressed in a CAS, whereas an off-the-shelf SAT solver is limited to expressions in Boolean logic.

Both SAT and CAS have their drawbacks when solving combinatorial problems. SAT solvers can perform scalable searches but lack the mathematical domain knowledge required to prune out symmetries in the search space. On the other hand, CASs have rich mathematical capabilities but lack scalability when dealing with enormous search spaces.
To combine the best of both worlds, we leverage \MC to dynamically provide mathematical context to the SAT solver to only enumerate non-isomorphic graphs in the search space.
Specifically, we use a CAS to generate blocking clauses that are passed to the SAT solver dynamically via a programmatic interface.
These clauses block the SAT solver from exploring symmetric branches of the search tree (see Section~\ref{sec:orderly}).
The SAT+CAS paradigm has been applied to solving hard combinatorial problems such as the Williamson conjecture~\cite{williamson}, Lam's problem~\cite{Lams}, the Erd\H{o}s-Faber-Lovász Conjecture~\cite{kirchweger2023sat}, the Kochen--Specker problem~\cite{li2023sat,kssms}, and integer factorization~\cite{Ajani2024}.
We show that SAT+CAS is orders of magnitude faster than a SAT-only approach without compromising the verifiability of the result.

\begin{table}
\centering
\begin{tabular}{cccc}
$R(p,q)$ & $p=3$ & $p=4$ & $p=5$ \\ \hline
$q=3$ & 6 & & \\
$q=4$ & 9 & 18 & \\
$q=5$ & 14 & 25 & 43--46 \\
$q=6$ & 18 & 36--40 & 59--85 \\
$q=7$ & 23 & 49--58 & 80--133 \\
$q=8$ & \textbf{28} & 59--79 & 101--193 \\
$q=9$ & \textbf{36} & 73--105 & 133--282 \\
\end{tabular}
\caption{Exact values and bounds for Ramsey numbers $R(p, q)$. The bolded values are the Ramsey numbers verified in this paper. Some values are excluded from the table since $R(p, q) = R(q, p)$.}
\label{tbl:ramsey}
\end{table}

\subsection{Ramsey Problems}
Ramsey's theorem states that for every $p$, $q \in \mathbb{N}$, there exists an $n \in \mathbb{N}$ such that every graph of order $n$ contains either a $p$-clique or an independent set of size $q$.
An $m$-clique is a complete subgraph of order $m$ and an independent set is a subset of mutually unconnected vertices.
The Ramsey problem is defined as finding the smallest integer $n$, denoted as $R(p, q)$, for some given $p$ and~$q$. 
A common and equivalent reformulation is as follows:
for every $p$, $q \in \mathbb{N}$, there exists an $n \in \mathbb{N}$ such that 
any red/blue coloring of the edges of the complete graph of order $n$, denoted $K_n$, contains a blue monochromatic $p$-clique or a red monochromatic $q$-clique. 
A Ramsey $(p,q)$-graph is a graph without a $p$-clique and without an independent set of size~$q$.  
Ramsey $(p,q;n)$-graphs and $(p,q;n;e)$-graphs are $(p,q)$-graphs on $n$ vertices and $(p,q)$-graphs on $n$ vertices with $e$ edges, respectively.
All graphs are assumed to be simple and undirected unless stated otherwise.

\subsection{Correctness of Results}

Correctness of results is a long-standing problem in the field of computer-assisted proofs, particularly for results that require extensive enumeration
that cannot be checked by hand.
Verification is of utmost importance, as it provides a formal guarantee that the result is correct. Without verification, one has to trust the correctness of the program, and this could allow undetected software or administrative errors to compromise the validity of the proof.
For example, recent work uncovered consistency issues in previous computational 
resolutions of Lam's problem, highlighting the difficulty of relying on search code for nonexistent results~\cite{Lams}. For prominent combinatorial problems such as Ramsey problems relying on extensive computations, formal verification is a way to increase trust in the results.

\begin{figure}
  \centering
  \includegraphics[scale=0.5]{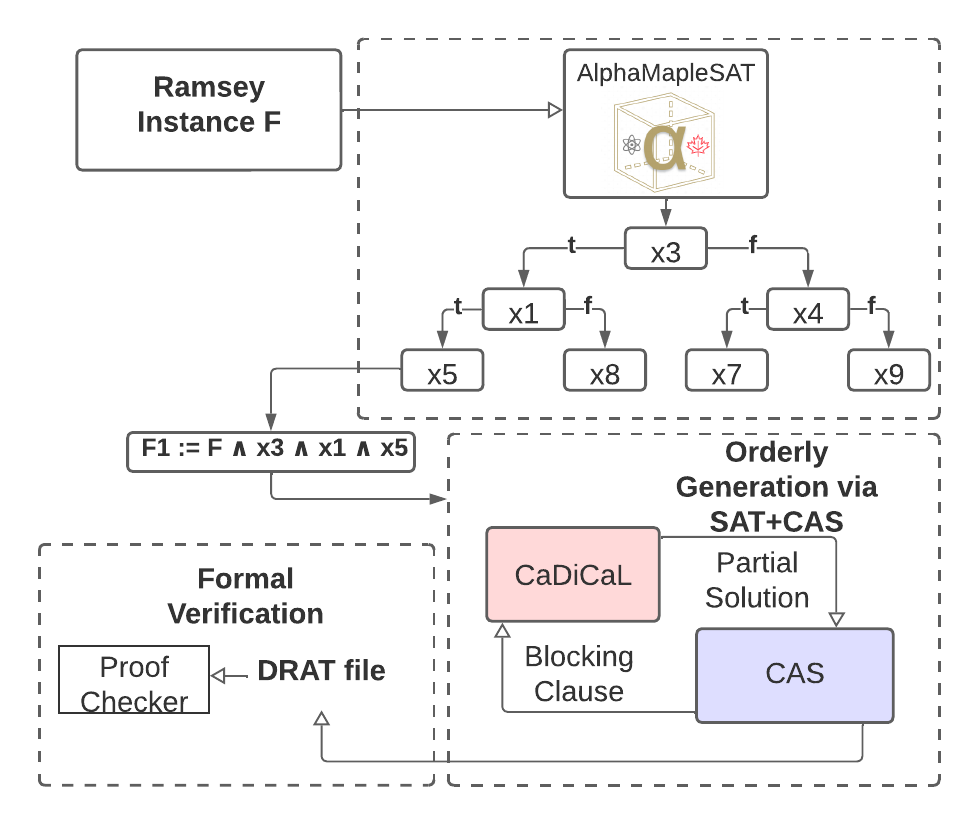}
  \caption{Flowchart of the parallelized tool. \AMS is used as the cubing solver, and \CaDiCaL + CAS is used as the conquering solver.}
  \label{fig:pipeline}
\end{figure}

\section{Previous Work}

\subsection{Classical Ramsey Numbers}

\citein{r45_formal} formally proved $R(4, 5) = 25$ using a SAT solver, verifying a result originally obtained in 1995 using computational search~\cite{r_45_mckay}.
Gauthier and Brown's approach combined an interactive theorem prover, a SAT solver, and gluing together generalizations of colored graphs as described below.
\emph{Graph gluing} refers to inserting an $n \times n$ adjacency matrix and an $m \times m$ adjacency matrix along the main diagonal of a larger empty matrix, to form an $(m+n) \times (m+n)$ adjacency matrix. 
The remaining off-diagonal entries are then filled in subject to some constraints (in this case the Ramsey constraints) in an exhaustive way.
A \emph{generalization} of a colored graph is a colored graph with some edges uncolored. 
A $(p,q)$-graph generalization with one uncolored edge represents the $(p,q)$-graph where the uncolored edge can be colored red or blue.
They constructed exact covers for sets of $(3, 5; m)$-graphs and $(4, 4; n)$-graphs, reducing the number of graph gluings required. 
Their method took over 2.5 years of CPU time, but they estimate it would have taken 44 years without generalizations. This represents a very recent advancement in the formal verification of results concerning Ramsey numbers, showcasing the potential of certifiable computational methods.

\citein{fujita2013scsat} used a soft-constraint approach to improve the lower bound of $R(4, 8)$ from 56 to 58. 
They introduced two types of soft constraints: zebra soft-constraints and unit soft-constraints. These could be iteratively removed, with their selection based on the number of conflicts relating to a soft-constraint.
This allowed for efficient propagation of edge assignments and significantly reduced the search space.

SAT Modulo Symmetries (SMS) is a framework developed for graph generation and enumeration~\cite{pysms}.
It leverages the SAT solver \CaDiCaL~\cite{Biere2024} with a dedicated symmetry propagator to check the canonicity of partial solutions. SMS is implemented using the IPASIR-UP interface~\cite{fazekas2023ipasir}, which allows the integration of user-defined symmetry propagators directly into the solving process, enabling efficient dynamic symmetry-breaking constraints.
This approach has been applied to verify smaller Ramsey numbers, such as $R(3,5)$ and $R(4,4)$~\cite{SMS2ramsey}, but has not yet been extended to larger instances. SMS uses a row-wise lexicographic ordering of the adjacency matrix for comparing graphs, whereas our method uses a column-wise ordering of edge variables above the diagonal---our method
requires the latter ordering because of a particular property not satisfied by the ordering used by SMS (see Property 2 in Section~5). In addition, SMS performs canonicity checks on partially defined graphs during the solving process, while our approach only performs canonicity checks on fully defined subgraphs (whenever an upper-left submatrix becomes fully known). We opted for SAT+CAS due to the availability of its verification capability, which is not currently provided in the SMS repository.

More recently, \citein{codelverified} introduced verified proof checking tools for the substitution redundancy (SR) proof system~\cite{sr1,sr2,sr3},
a powerful generalization of the propagation redundancy (PR)~\cite{pr} and resolution asymmetric tautology (RAT) proof systems~\cite{rat1,rat2,wetzler2014drat}.
Their work presents the first verified SR proof checker, implemented in the \textsc{Lean} theorem prover~\cite{deMoura2015},
and demonstrates SR's ability to produce significantly shorter proofs than RAT for certain problems.
They provide a concise 38-clause SR proof of $R(4,4) \leq 18$. Their experimental results show SR proofs are on average 99.6\% smaller than equivalent RAT proofs, while their verified checker performs comparably to existing fast PR checkers. Although no current state-of-the-art SAT solvers support SR reasoning, this work lays the groundwork for potential advancements in SAT solving techniques, offering a path to more powerful reasoning and shorter proofs for complex problems.

\subsection{Tri-color Ramsey Problems}

An $(r_1,\dotsc,r_k; n)$ Ramsey colouring is an edge $k$-coloring of $K_n$ with no monochromatic $K_{r_i}$ in color~$i$. The multicolor Ramsey number $R(r_1,\dotsc,r_k)$ is the smallest $n$ such that no such coloring exists.
\citein{codishresult} resolved the long-standing open problem $R(4,3,3)$ by showing that no $(4,3,3;30)$ coloring exists. They partitioned the search space into six $\langle a,b,c\rangle$-regular cases, where each vertex has fixed degrees in the three colors
\begin{gather*}
\langle 13, 8, 8\rangle,
\langle14, 8, 7\rangle,
\langle15, 7, 7\rangle, \\
\langle15, 8, 6\rangle,
\langle16, 7, 6\rangle,
\text{ or } \langle16, 8, 5\rangle.
\end{gather*}
All cases were proven unsatisfiable via SAT solving except $\langle13,8,8\rangle$. For that case, they generated all 3-colorings of $K_{13}$ without monochromatic triangles and attempted to extend each to $K_{30}$. The full computation required 128 CPU years, parallelized across 456 threads.

\subsection{Directed Ramsey Graphs}

A \emph{tournament} is a directed graph where for every pair of vertices, exactly one directed edge exists. A tournament is \emph{transitive} if for all vertices $u$, $v$, and $w$, the presence of edges $uv$ and $vw$ implies the existence of $uw$. The directed Ramsey number $R(k)$ is the smallest number of vertices such that every tournament on that many vertices contains a transitive subtournament of size~$k$.

\citein{heuleR7} improved the known bounds on $R(7)$ to 34 and 47 using SAT solving. For the upper bound, they encoded tournaments avoiding transitive subtournaments of size~7 and exhaustively searched all possibilities up to 46 vertices. For the lower bound, they applied a SAT solver to show no 33-vertex transitive-free tournament exists. Their approach combined constraint encodings with techniques such as self-subsuming resolution.

\subsection{Previous Work on \texorpdfstring{$\boldsymbol{R(3, 8)}$}{\textbf{R(3,8)}} and \texorpdfstring{$\boldsymbol{R(3, 9)}$}{\textbf{R(3,9)}}}

\citein{mckay1992value} computationally showed $R(3, 8) = 28$. Their method involved generating all graphs up to isomorphism on $20$--$22$ vertices without 3-cliques and without independent sets of size 7. They used a recursive generation procedure and the graph isomorphism tool \nauty\ to remove isomorphic graphs.

\citein{GRAVER1968125} made significant progress on $R(3, 9)$ before the problem was ultimately solved by \citein{grinstead1982ramsey}. Graver and Yackel proved $R(3, 9) \geq 35$ and showed a $(3, 9; 36)$-graph must be regular of degree 8 and must contain a $(3, 8; 27; 80)$-subgraph. Grinstead and Roberts computationally showed that no $(3, 8; 27; 80)$-graphs exist, thereby proving $R(3, 9) = 36$. Their method involved a series of lemmas on the structure of various subgraphs of $(3, 8; 27; 80)$-graphs and used graph gluing techniques combined with computational searches.

Since the previous approaches to $R(3, 8)$ and $R(3, 9)$ relied on uncertifiable computational methods for graph enumeration,
we therefore present certifiable proofs using the SAT+CAS paradigm.

\section{Encoding Ramsey Problems}
\label{sec:constraints}

We now discuss the encodings we used to translate the Ramsey problem into formulas in conjunctive normal form (CNF)\@.
In addition to the direct encoding of Ramsey graphs, we introduce encodings that further narrow down the search space by limiting symmetries, vertex degrees, and the number of edges in the graph.

\subsection{Encoding Ramsey Graphs}
The Ramsey problem is encoded for a predefined $n$, $p$, and~$q$ by deriving a Boolean formula in conjunctive normal form asserting the existence of a Ramsey $(p,q)$-graph of order~$n$.
The encoding enforces every $p$-clique to have at least one edge in the opposing (red) color and every $q$-clique to have at least one edge in the opposing (blue) color, i.e., 
\[ \bigwedge_{K_p \subseteq K_n} \bigvee_{e \in K_p} \neg e \quad\text{and}\quad \bigwedge_{K_q \subseteq K_n} \bigvee_{e \in K_q} e, \]
where the variable $e$ is assigned true~($\top$) when the corresponding edge is colored blue and is assigned false~($\bot$) when the corresponding edge is colored red.
A satisfying assignment of the encoding corresponds to finding a $(p,q)$-graph of order~$n$, and therefore $R(p,q)>n$.
Similarly, an unsatisfiable result means no such colorings exist for this particular~$n$, i.e., all colorings contain a blue $p$-clique or a red $q$-clique,
and therefore $R(p,q)\leq n$.

\subsection{Symmetry Breaking Constraints}

In order to break the symmetries of the problem, we first add static constraints enforcing a lexicographic ordering on rows of
the graph's adjacency matrix. After that, we use the dynamic symmetry breaking capabilities of a CAS to break the remaining symmetries. 

The partial static symmetry breaking constraints were developed by \citein{codish2019constraints}.
These enforce a lexicographical ordering on the rows of a graph's adjacency matrix and
block the solver from exploring certain symmetric portions of the search space before the CAS is called.
This is beneficial as there is an overhead associated with calling the CAS\@.

These clauses are constructed in the following manner:
for an adjacency matrix $A$ of a graph of order $n$, define $A_{i,j}$ as the $i$th row of A without columns $i$ and $j$.
The clauses enforce that $A_{i,j}$ is lexicographically equal or less than $A_{j,i}$ for all $1 \le i < j \le n$. 
These clauses introduce $O(n^3)$ auxiliary variables and clauses.
Based on our empirical evidence, these constraints provide a significant speed-up by breaking symmetries statically and were included in all instances.

\subsection{Cardinality Constraints}\label{sec:card}

Cardinality constraints are used to further reduce the search space by limiting both the degree of vertices and the number of edges in Ramsey graphs. Specifically, it is known that every vertex~$v$ of a Ramsey $(p,q;n)$-graph satisfies
\[ n - R(p,q - 1) \le \deg_b(v) \le R(p - 1,q) - 1 \]
where $\deg_b(v)$ is the number of blue edges on vertex~$v$~\cite{GRAVER1968125}. We also leverage theoretical results to restrict the number of edges allowed in the Ramsey graph when proving the value of $R(3,9)$ in Section~\ref{r39theory}.

We encoded these constraints using the totalizer encoding of \citein{b_b}.
The totalizer uses a binary tree to create relationships between auxiliary variables. Each node in the tree is assigned a value and a set of unique
variables. 
Suppose we wish to encode between $l$ and~$u$ of~$m$ variables are true.  We form a binary tree with
$m$ leaf nodes and associate each leaf node to one of these variables.
For a non-leaf node $r$ with children $a,b$, let $R= \{r_1,\dotsc,r_{m_0}\}$, $A=\{a_1,\dotsc,a_{m_1}\}$, and $B= \{b_1,\dotsc,b_{m_2}\}$ be the set of variables assigned to $r$, $a$ and $b$ respectively. 
The following conjunction of clauses is related to the node~$r$: 
\[ \bigwedge_{\substack{0\le\alpha\le m_1 \\ 0\le \beta \leq m_2 \\ 0\le \sigma \le {m_0} \\ \alpha + \beta =\sigma}}
(\neg a_\alpha \lor\neg b_\beta\lor r_\sigma) \land
(a_{\alpha+1} \lor b_{\beta+1} \lor \neg r_{\sigma+1})
\]
with $a_0$, $b_0$, and $r_0$ assigned true
and $a_{m_{1}+1}$, $b_{m_{2}+1}$, and $r_{m_0+1}$ assigned false.
These clauses ensure that the number of variables assigned true in $R$
is equal to the number of variables assigned true in $A\cup B$.
Finally, the clauses
$\bigwedge_{1\le i\le l}c_i$ and $\bigwedge_{u+1\le i\le m}\neg c_i$
specify that between $l$ and $u$ of our original $m$ variables are true
where $c_1,\dotsc,c_m$ denote the variables associated with the root node of the tree.
This encoding uses $O(m\log m)$ new variables and $O(mu)$ clauses after applying unit propagation and removing satisfied clauses.

\section{Orderly Generation Using SAT+CAS}\label{sec:orderly}

When searching for graphs using a SAT solver, the static symmetry breaking constraints introduced previously do not block all isomorphic copies---in fact, most symmetries remain in the search space.
Therefore, we use a symbolic computation method to block the remaining symmetries dynamically during the solving process.
Specifically, we implement an isomorph-free graph generation technique called orderly generation that was developed independently by \citein{read1978every} and \citein{faradvzev1978constructive}. 

We say an adjacency matrix $A_G$ of a graph $G$ is \emph{canonical} if every permutation of the graph's
vertices produces a matrix lexicographically greater than or equal to $A_G$,
where the lexicographical order is defined by concatenating the above-diagonal
entries of the columns of the adjacency matrix starting from the left.
In other words, the canonical matrix of a graph is the lexicographically-least (lex-least)
way of representing the graph's adjacency matrix.

An \emph{intermediate} matrix of $A_G$ is a square upper-left submatrix of $A_G$.  If $A_G$ is of order~$n$
then its intermediate matrix of order $n-1$ is said to be its \emph{parent},
and $A$ is said to be a descendant of its intermediate matrices.
The orderly generation method is based on the following two consequences of this definition of canonicity:
\begin{enumerate}
    \item[(1)] Every isomorphic class of graphs has exactly one canonical (lex-least) representative.
    \item[(2)] If a matrix is lex-least canonical, then its parent is also lex-least canonical.
\end{enumerate}

This definition of canonicity is particularly useful
as the contrapositive of the second property implies that if a matrix is not canonical,
then all of its descendants are also not canonical.
Therefore, any noncanonical intermediate matrix encountered during the search can be blocked immediately, as none of its descendants are canonical.

To implement the orderly generation algorithm in a SAT solver,
when the solver finds a partial assignment corresponding to an intermediate matrix, the canonicity of this matrix is determined by a canonicity-checking routine implemented in the CAS (Figure \ref{fig:pipeline}). The CAS explores permutations of the graph's vertices and evaluates whether any of these permutations result in a lexicographically smaller adjacency matrix---if so, the matrix is not canonical.
If the CAS finds a permutation that demonstrates the noncanonicity of the matrix,
then a blocking clause is learned which removes this matrix and all of its descendants from the search.
Otherwise, the matrix is canonical and the SAT solver proceeds as normal. 
When a matrix is noncanonical, the canonicity-checking routine provides a witness
(a permutation of the vertices producing a lex-smaller adjacency matrix),
which allows for the verification of \MC's result without trusting the CAS\@.

Even though $R(3,k)$ and $R(k,3)$ share the same value, in the context of the CNF encoding the $R(3,k)$ and $R(3,k)$ instances are different with $R(k,3)$ containing a much higher number of negative literals. An important observation in our experiments is that solving \( R(k,3) \) is consistently faster than solving \( R(3,k) \), despite these instances being equivalent due to symmetry. This behavior persists even when the CAS is turned off (see Table~\ref{tbl:sec6}).
We solved $R(3,k)$ and $R(k,3)$ for $k=8$ and~$9$ and included both runtimes in Section~\ref{sec:result}. 

\begin{table}
\centering
\begin{tabular}{c cc cc}
    & \multicolumn{2}{c}{\CaDiCaL + CAS} & \multicolumn{2}{c}{\CaDiCaL only} \\ 
$k$ & $R(3, k)$ & $R(k, 3)$ & $R(3, k)$ & $R(k, 3)$ \\ \hline
7   & 14.3 s& 8.2 s& 564.3 s& 220.7 s\\ 
8   & 112.1 h& 18.5 h& $>$ 7 days & $>$ 7 days \\
\end{tabular}
\caption{
Comparison of sequential runtime for instances $R(3,k)$ and $R(k,3)$ with $k=7$ and~$8$. ``\CaDiCaL + CAS'' indicates solutions using \CaDiCaL with CAS, while ``\CaDiCaL only'' uses \CaDiCaL without CAS\@. Cardinality constraints are excluded for $k=7$ to avoid making the instance too easy, but included for $k=8$. Experiments were conducted on Dual Xeon Gold 6226 processors running at 2.70 GHz.}
\label{tbl:sec6}
\end{table}

\section{Parallelization}\label{sec:c&c}

In order to scale our technique up to $k=9$, parallelization is used to reduce the wall clock time. We use a cube and conquer technique which splits the CNF instance into tens of thousands of subproblems and solves them in parallel.

\subsection{Cube and Conquer}
Cube and conquer~\cite{March} is a parallelization technique whereby a set of simpler instances are solved, and the aggregate result is equivalent to solving the original instance.
Initially developed to solve SAT instances arising from computing van der Waerden numbers~\cite{Ahmed2014},
many combinatorial problems have since
been attacked using this technique,
such as Lam's Problem~\cite{Lams},
the Boolean Pythagorean Triples Problem~\cite{heule_pyth_trip},
Schur number five~\cite{10.5555/3504035.3504843},
and the Kochen--Specker problem~\cite{li2023sat}.

Let $v_i$ be a variable appearing in the Boolean formula $F$. 
Then solving instances $F_1\coloneqq F \land v_i$ and $F_2\coloneqq F \land \neg v_i$ is equivalent to solving $F$, and
if either $F_1$ or $F_2$ is satisfiable then $F$ is satisfiable.
We say $v_i$ is the \emph{splitting} variable and consider $F_1$ and $F_2$ as subinstances of $F$.
If a subinstance is still hard, an unassigned variable (subject to some selection criteria) in the subinstance can be split on. 
This splitting can be repeated until some stopping criteria is reached.

When choosing the appropriate cubing solver, there are two main factors to consider: one is the time it takes to generate all cubes,
and the other is the quality of the cubes, which can be measured by the time required to solve the hardest cube.
Two cubing solvers we tried are \textsc{march\_cu}~\cite{March} and \AMS~\cite{alphamaplesat}.
We use \AMS as it generates a large set of cubes faster without compromising the quality of the cubes.

\AMS introduces a novel approach to cube challenging combinatorial instances. At its core, \AMS employs a Monte Carlo Tree Search (MCTS) based lookahead cubing technique, setting it apart from traditional cube-and-conquer solvers. This method allows for a deeper heuristic search to identify effective cubes while maintaining low computational costs. 

We define eliminated variables to be variables that are either assigned a $\top$/$\bot$ value or propagated to be $\top$/$\bot$ by a simplification solver 
(\CaDiCaL with orderly generation). To determine when to stop cubing the instance further, we used the number of eliminated variables as the stopping criteria. If the number of eliminated variables is greater than a predefined threshold, the cubing stops for this subinstance. Given the original formula $\phi$, each time \AMS chooses an edge variable $x$ to split on, the formula $\phi \land x$ and $\phi \land \lnot x$ are generated and simplified using \CaDiCaL + CAS, so that \AMS can choose the next splitting variable based on the simplified formula, since this contains CAS-derived clauses and therefore blocking clauses from orderly generation are incorporated during the cubing process.

Following cubing, the cubes are ``conquered'', i.e., solved by a SAT+CAS solver.
Ideally, the solver can solve all cubes.
However, when our solver could not solve an instance without producing a proof file larger than 7 GiB, the instance was passed back to \AMS to be cubed again.
7 GiB was chosen as the maximum file size because we found that the DRAT-trim proof checker~\cite{wetzler2014drat} could verify such proofs without exceeding our memory limit of 4~GiB\@.
This process of alternating cubing and conquering continues for difficult instances until all cubes can be solved with proof files smaller than 7~GiB\@. To solve cubes efficiently across multiple CPUs, we utilized Python's multiprocessing library.

\begin{table*}[ht]
\centering
\begin{tabular}{c c c c c c c c c}
\textbf{Instance} & \textbf{Cubing Time} & \textbf{Simplification Time} & \textbf{Solving Time} & \textbf{Verification Time} & \textbf{Wall Clock Time} \\ \hline
$R(8,3)$  & \phantom01,360 s & \phantom01,217 s & \phantom019,811 s & \phantom022,328 s & \phantom08 hrs   \\ 
$R(9,3)$  &         15,530 s &         42,482 s &         697,575 s &         473,874 s &         26 hrs   \\ 
\end{tabular}
\caption{Summary of experimental results for solving \( R(8,3) \) and \( R(9,3) \) in parallel.}
\label{tbl:results}
\end{table*}

\section{Theory of \texorpdfstring{$\boldsymbol{R(3,9)}$ and $\boldsymbol{R(9,3)}$}{Theory of R(3,9) and R(9,3)}}\label{r39theory}
Directly applying a SAT+CAS solver to the $R(9,3)$ problem is difficult.  
As the problem is significantly harder than the $R(8,3)$ problem, we leverage theoretical results to reformulate the problem.
\citein{grinstead1982ramsey} combined theory and computation to show that $R(3,9)=36$, leveraging theoretical results from \citein{GRAVER1968125}. This section highlights key theoretical results from Graver and Yackel's paper. 
The following lemma, from Section~3 of \citein{GRAVER1968125}, forms the basis of the original proof of $R(3,9)=36$.
We also use it in our proof and verification, since the lemma does not rely on computation and can be proved mathematically.

\begin{lemma}\label{lem:3-9-36}
    A $(3,9;36)$-graph contains a $(3,8;27;80)$-graph. A $(9,3;36)$-graph contains a $(8,3;27;271)$-graph.
\end{lemma}


Graver and Yackel improved the bounds on many Ramsey numbers and on the minimum number of edges in Ramsey graphs. They constructed a $(3,9)$-graph on 35 vertices, thereby showing $R(3,9)>35$.
Specifically for $R(3,9)$, Grinstead and Roberts derived a sequence of lemmas on the structures of various subgraphs of $(3,8;27;80)$-graphs. 
They computationally proved structures in $(3,8;27;80)$-graphs cannot exist in order to show $R(3,9)\le36$ via Lemma~\ref{lem:3-9-36}.
They estimated $5 \times 10^{10}$ machine operations and $2.5 \times 10^4$ seconds of computation,
but note the time could be improved using machines with more efficient bitstring operations as
computations were performed on a Honeywell Level 66 computer. The second part of the lemma is derived from the fact that $\Comb{27}{2} - 80 = 271$, thus a $(9,3;36)$-graph contains a $(8,3;27;271)$-graph.

We apply a parallelized SAT+CAS tool to a CNF encoding asserting the existence of a $(8,3;27;271)$-graph.
We obtained an UNSAT result (see Section~\ref{sec:r39}), therefore showing that no $(8,3;27;271)$-graphs exist.
By Lemma~\ref{lem:3-9-36}, this implies $R(9,3)\leq36$,
and since a $(3,9;35)$-graph exists, this implies $R(9,3) = R(3,9)=36$.
In doing so, we do not rely on the method employed by the authors of the original proof to show the
nonexistence of $(3,8;27;80)$-graphs, 
since they rely on unverified computational results.
To simplify our encoding, we do not rely
on any additional theoretical results on $(3,8;27;80)$-graphs or $(8,3;27;271)$-graphs, aside from the vertex degree constraints mentioned in Section~\ref{sec:card}.

\section{Results and Verification}\label{sec:result}

\subsection{Experimental Setup}

We performed both SAT+CAS solving and verification in parallel by integrating \AMS with the \MC tool.
The \( R(8,3) \) result was obtained on a cluster of Dual Xeon Gold 6226 processors @ 2.7 GHz and used 24 CPUs,
while the \( R(9,3) \) result was computed on a cluster of Dual AMD Epyc 7713 CPUs @ 2.0 GHz and used 128 CPUs.

By default, \MC uses a pseudo-canonical check and stops the canonical check
early if the CAS takes too long to determine canonicity.
To optimize solving performance, we enabled full canonicity checking and  
this adjustment resulted in a 2$\times$ improvement in solving time.
Even after enabling a full canonical check, a minority of the total solving time is spent in the CAS---
for $R(8,3)$, the CAS accounts for 7\% of the total solving time, while for $R(9,3)$, it accounts for 38\%.
Although a full canonicity check is more expensive, it generates more symmetry-blocking clauses
and ultimately enhances the solver's efficiency for this problem.

Table~\ref{tbl:results} includes ``simplification time'' for our cube-and-conquer tool. During cubing, each variable split triggers a brief {\CaDiCaL}+CAS simplification (10,000 conflicts), with the simplified instance passed back to AMS\@. This tracks eliminated variables and crucially allows AMS to leverage CAS-derived clauses for balanced cube generation.

\subsection{Solving \texorpdfstring{\( \boldsymbol{R(3,8)} \) and \( \boldsymbol{R(8,3)} \)}{\textbf{R(3,8) and R(8,3)}}}

The Ramsey number \( R(3,8) = 28 \) was confirmed by obtaining an UNSAT result for the encoding asserting the existence of a 28-vertex \( (8,3) \)-graph and a SAT result in 288 seconds for the encoding asserting the existence of a 27-vertex \( (8,3) \)-graph. Solving \( R(8,3) \) on 28 vertices can be done sequentially (as shown in Table~\ref{tbl:sec6}) or in parallel.

For $R(8,3)$, we used the cardinality constraints from Section~\ref{sec:card}, implying
that the degree of each vertex is between 20 and~22.
Cubing was performed until 120 edge variables were eliminated. If a cube was not solved after producing a 7~GiB certificate, further cubing was applied, eliminating an additional 40 variables.

A total of 41 cubes were generated, each returning UNSAT\@. With parallelization, the problem was solved in approximately 8 hours of wall clock time and the combined certificate files amounted to 5.8 GiB\@. We applied the same techniques to solve $R(3,8)$, where each vertex has a degree between 5 and~7. We observed that solving the $R(8,3)$ instance is about 6$\times$ faster than solving $R(3,8)$ sequentially.

\subsection{Solving \texorpdfstring{\( \boldsymbol{R(3,9)} \) and \( \boldsymbol{R(9,3)} \)}{\textbf{R(3,9) and R(9,3)}}}\label{sec:r39}

Solving \( R(9,3) \) corresponds to solving the \( R(8,3;27;271) \) problem, as mentioned in Section~\ref{r39theory}. This instance must be solved using the parallelized \MC tool, as \( R(8,3;27;271) \) is much more challenging to solve sequentially in a reasonable amount of wall clock time.

Cardinality constraints were applied similarly, enforcing that each vertex must have a degree between 19 and 22, and that the graph must contain exactly 271 edges. Cubing continued until 100 edge variables were eliminated. If a cube was not solved after producing a 7~GiB certificate, further cubing eliminated an additional 40 variables.

A total of 2,486 cubes were generated, each returning UNSAT\@. With parallelization, the problem was solved in approximately 26 hours of wall clock time and the combined proof files amounted to 289~GiB\@. Similarly, $R(3,9)$ is solved using the same cubing cutoff criteria. We witnessed that solving the $R(9,3)$ instance is about 6$\times$ faster than solving $R(3,9)$.

\subsection{Verification of Results}

The SAT+CAS approach produces verifiable certificates, enabling an independent third party to confirm the solver's results. These certificates ensure that the SAT solver's search is exhaustive and that the learned clauses provided by the CAS are correct. As a result, only the correctness of the proof verifier—a relatively simple piece of software—needs to be trusted, rather than the SAT solver or the CAS itself.

Verification was performed using the DRAT-trim proof checker~\cite{wetzler2014drat} modified to support clauses specified to be trusted. Clauses derived from the CAS were prefixed with the character `\texttt{t}' to mark them as trusted, and they were verified separately. A Python script applied the witness permutations to confirm that each noncanonical adjacency matrix produced a lexicographically smaller matrix, verifying the correctness of the CAS-derived clauses. Since the script verifies that each CAS-derived clause in the DRAT proof blocks only noncanonical partial assignments, we do not rely on the correctness of the somewhat complicated canonical check in the CAS\@.

When using the cube-and-conquer approach, it is critical to ensure that the generated cubes collectively partition the search space. Completeness was verified by recursively checking that for any literal \( x \) forming a cube \( \phi \land x \), all extensions of the complementary literal \( \phi \land \neg x \) were covered by the set of generated cubes. This process confirmed that \(R(8,3) = R(3,8) = 28 \) and \(R(9,3) = R(3,9) = 36 \). In addition, we ran $R(8,3)$ on the same machine sequentially and obtained a runtime of 66,504 seconds (see Table~\ref{tbl:results}), however, the total CPU time combining cubing, simplification, and solving was only 22,388 seconds. Therefore, cube-and-conquer achieved a reduction in total CPU time compared to sequential solving. This mirrors the efficiency gains observed in the Boolean Pythagorean Triples problem~\cite{heule_pyth_trip}, where cube-and-conquer reduced the total solving time from 2,125 days to 2 days using 800 cores, effectively reducing the computational effort to 1,600 CPU days. 

While our approach produces certificates of completeness for exhaustive search procedures, these certificates do not constitute formal proofs of the corresponding Ramsey numbers. Rather, they serve as computational validation of the most resource-intensive components of the argument, i.e., the parts which in practice cannot be checked by hand. To enable full formal verification of a Ramsey number within our pipeline, two key elements remain unverified: the correctness of the SAT encoding and the correctness of the custom script used to validate clauses generated by the CAS.

\section{Conclusion}
Using SAT+CAS, we significantly improve the efficiency of a SAT solver on Ramsey problems and
provide the first independently-checkable proof of the result $R(3,8)=28$ of McKay and Min from 1992.
We verify that $R(3,9)\le36$ (and therefore $R(3,9) = 36$)
with less dependency on theoretical results than previous approaches.  This
result was first shown by Grinstead and Roberts in 1982 but was never before verified
by independently checkable certificates.

SAT+CAS has been demonstrated to be an effective problem-solving and verifying tool for Ramsey problems.
When combined with previously known domain knowledge about Ramsey problems,
the search space can be reduced and thus the effectiveness of the SAT+CAS method is improved. 
More difficult Ramsey problems often have larger encodings and require an excessive amount of memory. 
Thus, reducing problems using domain knowledge has the additional benefit of reducing the encoding size.
Future applications of SAT+CAS to Ramsey problems include the verification of all known Ramsey numbers and determining the values of
$R(3,10)$, $R(4,6)$, or $R(5,5)$, which remain open problems.
While our certificates are independently checkable and ensure correctness of each individual SAT instance, a fully formal proof of the main theorem would require formalizing the correctness of the CNF encoding, verifying the checker used to validate UNSAT claims, verifying the correctness of Lemma~\ref{lem:3-9-36}, and connecting the SAT results to the formal definition of the Ramsey number in a proof assistant such as \textsc{Lean} or \textsc{Coq}. We leave this comprehensive formalization to future work.

\bibliographystyle{named}
\bibliography{ijcai25}

\end{document}